# Designing Applications in a Hybrid Cloud

**Evgeny Nikulchev**

Moscow Technological Institute, Moscow, Russia

**Evgeniy Pluzhnik**

Moscow Technological Institute, Moscow, Russia

**Dmitry Biryukov**

Moscow State Technical University of Radio Engineering, Electronics
and Automatics, Moscow, Russia

**Oleg Lukyanchikov**

Moscow State Technical University of Radio Engineering, Electronics
and Automatics, Moscow, Russia

**Abstract**

Designing applications for hybrid cloud has many features, including dynamic virtualization management and route switching. This makes it impossible to evaluate the query and hence the optimal distribution of data. In this paper, we formulate the main challenges of designing and simulation, offer installation for processing.

## 1 Introduction

Currently cloud technologies are evolving rapidly. Cloud computing is now a household term introduced by marketers for the sale of information services. Cloud system is a typical client-server system that is focused on servicing the large number of customers. To ensure the required level of service information, users need to be constantly monitored while managing the system, as the number of users and their activity is constantly changing. To do this, the system must be flexible, scalable, and secure.

Hybrid cloud systems are becoming much more popular, as they allow to partially solve the basic problem of cloud security. Hybrid infrastructure has many positive aspects of cloud computing: scalability and virtualization, but also the safety and security of the data [1] due to its distribution. Under hybrid cloud infrastructure we understand databases, distributed private and public clouds, as well as communication channels and programs, providing for interaction between the components of the database [2].

To build high-quality cloud-based system, you must follow some basic principles for creating distributed information systems [3]:
– the introduction of the principles of dynamic control;
– the use of special technologies for the development;
– the need to evaluate computational costs.

For dynamic management of cloud infrastructure prompted to implement feedback. Note that the idea of feedback in computer systems was developed [3]. In particular, in cloud technologies different ways of constructing a system of dynamic equations are offered [4]. Providing feedback is possible by means of hypervisors used to build modern cloud systems and the software in cloud-based systems, if such opportunities were provided by developers. The introduction of a feedback with one hand allows you to make corrective action, with another — is part of the computing resources and communication channels. Therefore it is necessary for optimal control problems, where the quality criteria can minimize the processing time or resource constraints.

On the basis of these feedbacks decisions are made to redesign the entire system. Often this occurs when changing the security or reliability of the system, as well as in cases where to increase system performance at the expense of computational resources is not possible or not economically viable. In result almost new system is developed, often from scratch by new division of programmers. Obviously, this is     labor-intensive and not profitable, therefore, is the urgent task of creating programming technologies that would provide high flexibility to the systems, to modify them, if necessary, without much effort.

## 2 Cloud system control based on feedback

One of the tasks of cloud system management is the task of load balancing of computing resources. If given insufficient computing resources, it is not ensured the required level of service. If you have selected an excessive amount, it is not economically viable to buy or rent extra equipment. To select the optimal number of computing resources to provide the required level of service, feedback is implemented.

Today, any large-scale cloud system is based on a virtual basis, which provides a convenient way to change the amount of computational resources depend-ing on needs, which are calculated on the basis of feedback. Almost all virtualization systems provide feedback, so that the main computational parameters can be monitored: CPU usage, RAM, hard disk and network.

The author carried out experiments on a test stand (Fig. 1), described in [3, 5], which uses virtualization via VMware ESXi. Virtualization environment is implemented VMware ESXi, installed using flash memory. Configured virtual switch Cisco Nexus 1000 and deployed 4 virtual machines on the physical disk server.

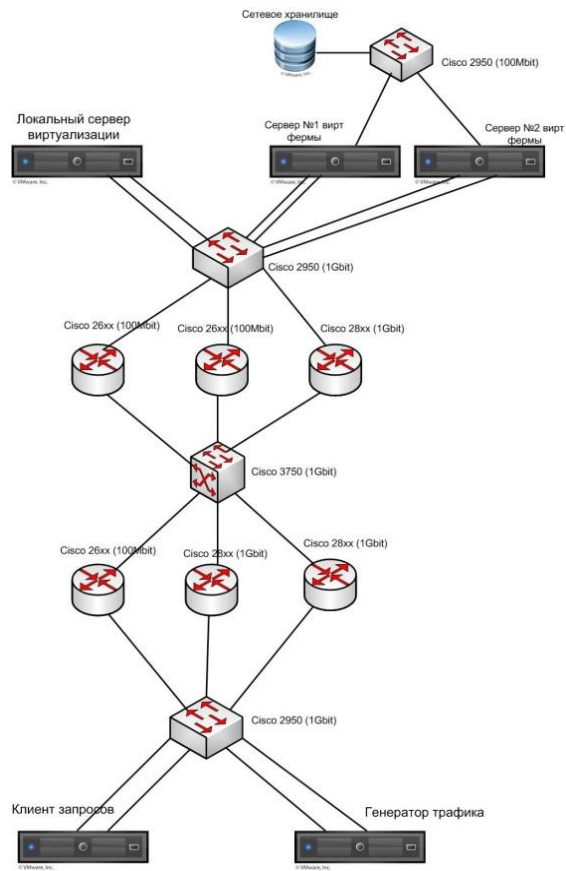

Fig.1: Experimental installation

For organizations using cloud product family VMware vCloud, cloud computing allows you to organize at all levels. To create a cloud in the experimental stand on two servers SunFire hosts created VMware ESXi, established management system VCenter, installed VMware vCloud Director.

The presence of more than 15 physical Cisco switches and routers 29 Series 26 and Series 28, as well as the virtual switches Nexus, the functional network equipment, the use of dynamic routing protocols, technologies Vlan, trunk, QoS and other allows for a variety of network diagrams.

VMware ESXi independently distributes the load of virtual machines, but also contains a number of add-ons that allow doing this manually by developers or admi-nistrators. To automate the management and development of own methods of load balancing there is used the functionality of the VMware libraries or the PowerCLI console. PowerCLI is an extension for Windows Powershell, which adds over 400 new commands for managing virtual infrastructure, including Cloud. The results of running the PowerCLI commands return results in the format of objects .NET, making it easy to develop tools in C# with a library .NET Framework.

To provide system virtualization management of VMware ESXi the next set of commands is used (detailed description of all commands PowerCLI can be found in the reference [6]):

– Connect-VIServer — connects to the virtualization system, after which you can perform the rest of the commands;
– Get-VMHost — gets information about the host, including its computational characteristics;
– Get-VM — gets the list of virtual machines and their computational characteristics;
Get-Stat — gets the downloaded history of computing resources;
– Set-VMResourceConfiguration — changes the limits and priorities for the distribution of computing resources of the host.

The organization of the system architecture does not always provide the required service for users, even with great computational resources. To improve the efficiency of the system and its security the distribution of data and computations are often used. For making decisions about the optimal allocation, it is necessary to collect a wider range of parameters from the system, which should be provided by implementing feedback on the system design phase. With feedback system it is possible to perform the profiling of the system with real workloads, and also to provide statistical data for the choice of strategy of development of the project cloud. For example, experiments (described in [7]) in which the feedback parameter is taken as the execution time of a relational query, according to which the decision to transfer part of the database in the private part of the cloud, with the aim of improving the safety and performance of their systems. In addition to the execution time of queries, developers can include parameters, for example, responsible for the geographical distribution of users, the statistics on user activity in time and statistics on the frequency of request for varied information. These settings will allow giving recommendations on the distribution system for greater efficiency. However, adding this functionality to developers is labor-intensive, so it is important to develop programming techniques for cloud systems, providing statistical data and facilitating easy to change the architecture of the system.

The organization of the system architecture does not always provide the required service for users, even with high amount of computational resources; in addition to this, there are also requirements for security and reliability. To make a decision about system's redesign and its distribution wide range of parameters is needed, for that there was implemented feedback system. Settings from the software that can make recommendations for redesign of the system include:

– the execution time of the operations (queries, functions of services, etc.);
– the frequency of the data request;
– geographical location of users and their activity.

For example, in [3] experiments are described in which time-based execution of queries to the database, the possibility of migration to a hybrid cloud infrastructure, with the aim of enhancing the security of the system. And experiments have shown that for public users there will be a significant loss in performance.

## 3 Agent-relational mapping technology

In the process of functioning of the information systems associated with the processing of big data, there are often situations that require restructuring of data storage structures and architecture of the system itself. Changing the structure, of course, would entail a change in the software implementation of all the data requests, and as a result, the reprogramming. In these conditions software middleware is required, taking on the task of changing the structure and physical location of the data.

There are several main types of middleware [8]:
– DOT (*distributed object technology*);
– DB oriented technologies (*database access middleware)*;
– RPC (*remote procedure call*);
– MOM (*message-oriented middleware*).

The developer is usually tied to one or a couple of middleware technologies, building on them the entire system. All remote interaction between the objects and the exchange of data in the system is carried out on the selected concept. This limits the abilities of the system; also often ineffective solution is selected to implement functionality not fitting the technology, because it is less labor-intensive. To solve these problems, a new unified middleware technology for programming distributed systems is advised, combining the concept of middleware technologies.

Agent-relational mapping (ARM) — displays the relational model to the object model, elements of which have the ability to remotely interact with each other [9].

The technology, which realizes the ARM, is joint use of the concept of remote access and ORM, and additional functionality set forth in the following three theoretical positions.

1. Remote interaction between objects and sending commands is relational commands select tool: change (*update*), add (*insert*), delete (*delete*).
2. The presence of transactions when performing operations.
3. Storing information about remote objects.

As a result, ARM technology has an interface of ORM technology, but allows you to perform operations not only to the database, but also to remote objects located on other computing nodes. Main features ARC described in Table 1.

Table 1. ARM technology features.

| ARM features | Technologies |
|---|---|
| SQL processing is hidden: all requests are generated and executed independently | DOT (*distributed object technology*) |
| Provides functions for interaction with other similar objects | DB oriented technologies (*database access middleware*) |
| Remote function calls on DBMS and separated nodes | RPC (*remote procedure call*) |
| Indicates the end of the operations *update*, *insert*, *delete* | MOM (*message-oriented middleware*). |
| Provides synchronization of all objects in the software system. | |

A prototype of the ARM technology, which includes the service «ArNotifyService» and used to provide network communication is shown in Fig. 2. Library component «ArLib» (partly shown in Fig. 3) receives ARM opportunities.

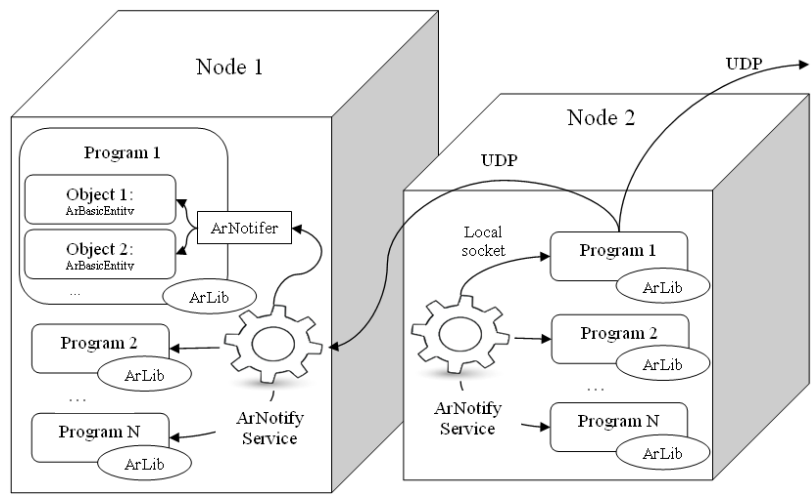

Fig. 2. Principle of communication service «ArNotifyService»

«DbProcess» module provides connection to the database and querying. «QueryOptions» module provides storage requests in a special structure, as well as generates necessary inquiries with the sample filters and sorting. «DbEntity», «DbEntityLink», «DbEntityView» modules provide direct agent-relational mapping. «ArNotifer» module provides interaction between objects.

«ArNotifer» class is a singleton, a copy of which is in the application, can be created only one. «ArNotifer» object performs direct transmission and reception of messages containing information about the sender and recipient, serialized object. Message exchange occurs according to the scheme shown in Figure 1, using 3-tier addressing: host (ip or url), app (pid number) and the object (the object name and identifier)

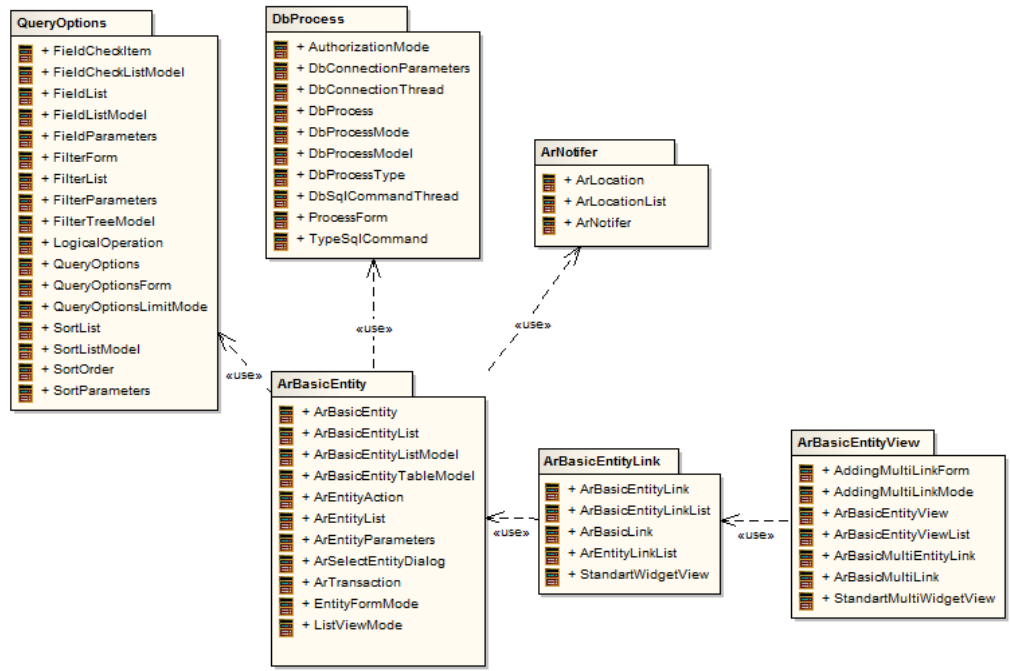

Fig. 2. «ArLib» library classes

## 4 Conclusion

The article describes the basic principles of control and management of cloud systems. Effective management is possible only with feedback from the cloud, which can be achieved by the hypervisor or the specialized capabilities of the software systems, if they were provided in advance by developers. Additional feedback, especially on the software, allows us to give more information to effectively manage the balancing of resources and to evaluate the effectiveness of the system. Additional feedback, on the one hand, leads to the corrective action, on the other — takes a part of computing resources and communication channels, so that the required optimal control problem, where quality criteria can minimize the processing time or resource constraints. If increasing computational resources is not economically viable (based on the assessment of the effectiveness of the system), decision to redesign the entire system is made. To change the architecture, if necessary, agent-related mapping technology is advised, which will provide cloud-based system with necessary flexibility.